\documentstyle[12pt,aasms4,psfig]{article}

\slugcomment{Astrophysical Journal, February 2000, 530, 618 (plus erratum)}
 
\begin{document}
 
\title{The Scatter in the Relationship between
Redshift and the Radio-to-Submm Spectral Index}
 
\author{C. L. Carilli}
\affil{National Radio Astronomy Observatory, P.O. Box O, Socorro, NM,
87801, USA \\
Max Planck Institute for Radio Astronomy, Auf dem H\"ugel 69, Bonn,
Germany, D53121} 
\authoremail{ccarilli@nrao.edu}
\author{Min S. Yun}
\affil{National Radio Astronomy Observatory, P.O. Box O, Socorro, NM,
87801, USA \\}
\authoremail{myun@nrao.edu}
 
\begin{abstract}

We derive the scatter in the relationship between redshift and
radio-to-submm spectral index, $\alpha^{350}_{1.4}$, using the
observed spectral energy distributions of 17 low redshift star forming
galaxies. A mean galaxy model is derived, along with the rms scatter
in $\alpha^{350}_{1.4}$. The scatter is roughly constant with
redshift. Constant rms scatter, combined with the flattening of the mean
$\alpha^{350}_{1.4}$ -- $z$ relationship with increasing redshift,
leads to increasing uncertainty for redshift estimates at high
redshifts. Normalizing
by the dust temperature in the manner proposed by Blain decreases the
scatter in $\alpha^{350}_{1.4}$ for most of the sample, but does not
remove outliers, and free-free absorption at rest
frequencies above 1.4 GHz is not likely to be a dominant cause for
scatter in the $\alpha^{350}_{1.4}$ -- $z$ relationship. We re-derive
the cumulative redshift distribution of the 14 field galaxies in a
recent submm and radio source sample of Smail et al..  The most likely
median redshift for the distribution is 2.7, with a conservative lower
limit of $z = 2$, as was also found by Smail et al. based on the original
$\alpha^{350}_{1.4}$ -- $z$ models. The normalization and shape of the
redshift distribution for the faint submm sources are
consistent with those expected for forming elliptical galaxies.

\end{abstract}
 

\section {Introduction}

The sharp rise of observed flux density, $S_\nu$, with increasing
frequency, $\nu$, in the Rayleigh-Jeans portion of the grey-body
spectrum for thermal dust emission from star forming galaxies ($S_\nu
\propto \nu^{3 - 4}$), leads to a dramatic negative $K$-correction for
observed flux density with increasing redshift. Submm surveys thereby
provide a uniquely {\it distance independent} sample of objects in the
universe, meaning the observed submm flux density of an object of
given intrinsic luminosity is roughly constant for $z$ between 1 and
7 (Blain \& Longair 1993,
Hughes \& Dunlop 1998).  Sensitive observations at submm wavelengths
are revealing what may be a population of active star forming galaxies
at high redshift which are unseen in deep optical surveys due to dust
obscuration (Smail, Ivison, and Blain 1997, Barger et al. 
1998, Hughes et al. 1998, Barger et al. 1999a, Blain et
al. 1999a, Eales et al. 1999, Bertoldi et
al. 1999). Current models suggest that this population may represent
the formation of elliptical galaxies and galactic bulges at $z$
between 2 and 5, constituting about half of the total amount of cosmic
star formation from the big bang to the present (Barger et al. 1999b,
1999c, Tan, Silk, \& Balland 1999, Blain et al. 1999b, Trentham,
Blain, and Goldader 1999, Lilly et al. 1999, Smail et al. 1999c).

The faint submm source counts suggest that a significant revision is
needed to the optically derived star formation history of the universe
(Steidel et al. 1999), with the addition of a population of highly
reddened active star forming galaxies not seen in deep optical surveys
(Chapman et al. 1999).  However, this conclusion hinges on the
currently unknown redshift distribution of the faint submm sources.
Follow-up observations have shown that most faint submm sources are
associated with very faint optical ($R\ge25$) and near IR ($K\ge21$)
sources (Smail et al. 1999a), in which case obtaining reliable optical
redshifts is difficult. Moreover, at these faint levels confusion in
optical fields can be severe. At the limit of the Hubble
Deep Field one expects about 5 sources in a typical faint submm source error
circle of 3$''$ radius (Blain et al. 1999c, Downes et al. 1999, Hughes
et al. 1998). For comparison, the probability of detecting a radio
source with $\rm S_{1.4} \ge 50 \mu Jy$ in this same region is 0.04
(Richards 1999). Even crude estimates of source redshifts that do not
rely on identification of the sources in the optical or near IR are
fundamental to our understanding of the faint submm source population.

We recently proposed the technique of using the
radio-to-submm spectral index as a redshift indicator
(Carilli and Yun 1999).  This technique
is based on the universal radio-to-far infrared (FIR) correlation for
star forming galaxies (Condon 1992), with the extrapolation that the
spectral shapes may be similar enough to be able to differentiate
between low and high redshift objects. Even a
gross indication of the source redshift distribution provides critical
constraints on models of the star formation history of the universe
using submm source counts (Blain et al. 1999b).  Carilli and Yun
(1999) presented empirical models for the relationship between $z$ and
the observed spectral index between 1.4 and 350 GHz,
$\alpha^{350}_{1.4}$, based on the spectral energy distributions
(SEDs) of Arp 220 and M82, plus semi-analytic models based on the
equations in Condon (1992), assuming a dust spectrum comparable to
that of M82 (G\"usten et al. 1992).  We found that these simple models
were in reasonable agreement with the few spectroscopically measured
redshifts for high $z$ submm sources, and we discussed possible
reasons for departure of sources from this relationship, including the
presence of a radio loud AGN, and cooling of the synchrotron emitting
relativistic electrons due to inverse Compton scattering off the
microwave background radiation.  A subsequent paper by Blain (1999)
showed that very low dust temperatures ($\rm T_{D} < 30 K$) can also
lead to departures from the standard relations.

In this paper we take a complimentary, empirical,
approach to the analytic approach
of Blain (1999), by looking at the mean and scatter in the 
$\alpha^{350}_{1.4}$ -- $z$ relation using the observed SEDs of 
17 low $z$ star forming galaxies. 
Note that Blain (1999) considered the logarithm of the flux
density ratio between observed frequencies of
350 and 1.4 GHz while we use the spectral
index. These quantities are simply related by a scale factor: 
log[${350}\over{1.4}$] = 2.4. We use $H_o$ = 75 km s$^{-1}$
Mpc$^{-1}$ and q$_o$ = 0.5.

\section{Analysis}

\subsection{Scatter in the $\alpha^{350}_{1.4}$ -- $z$ Relation}

Most of the galaxies in our study come from the recent submm study of
Lisenfeld, Isaak, \& Hills (1999), which is a subsample of bright
IRAS galaxies observed in CO(1--0) by Sanders, Scoville, \& Soifer
(1991). Lisenfeld 
et al. detected 14 of 19 galaxies at 350 GHz using SCUBA on the
JCMT. As discussed in Lisenfeld et al., these galaxies span a wide
range in IR luminosity, from luminous ($\rm L_{FIR} \approx  
10^{10} L_\odot$), to ultraluminous ($\rm L_{FIR} \approx
10^{12} L_\odot$), and in infrared-to-blue luminosity ratio, 
and should be representative of the star forming
galaxy population, although the fact that these are IRAS selected
galaxies could lead to a bias toward warm dust. 
We have augmented this sample with three other star
forming galaxies with well sampled SEDs from 1.4 GHz to 25000 GHz
(M82, IRAS 05189$-$2524, and UGC 5101). The sources are listed in column
1 of Table 1. 

The process of deriving the redshift evolution of $\alpha^{350}_{1.4}$
entails fitting accurate polynomials to the observed SEDs at cm
through IR wavelengths, then using these polynomials to predict the
change in $\alpha^{350}_{1.4}$ with redshift (Carilli and Yun 1999).
Examples of the SEDs, plus polynomial fits,
for four of the galaxies are shown in Figure 1. 
The fitting process is required to interpolate the discretely sampled
data points to unsampled spectral regions. As with any interpolation
process, errors will occur in unsampled regions, and the errors will
increase as the sampling decreases. This is particularly true for the
six sources in the sample with only one mm or submm measurement (NGC
3110, 5135, 5256, 5653, 5936, Zw049). Such errors are difficult to
quantify. We have performed the analysis excluding these sources, and
find no substantial changes in the results.


Table 1 lists the spectral
index between 0.33 and 1.4 GHz (column 2), 1.4 and 5 GHz (column 3),
and 350 GHz and 1.4 GHz (column 4), the FIR luminosities (column 5),
and the radio-to-FIR ratio (column 6), as quantified in the $q$
parameter defined by Condon (1992):~ $q\equiv$ log [${S_{100} + 2.6
\times S_{60}}\over{3 \times S_{1.4}}$], where $S_{1.4}$ is the flux
density at 1.4 GHz in Jy, and $S_{60}$ and $S_{100}$ are the flux
densities at 60$\mu$m and 100 $\mu$m, in Jy, respectively.  The
sources are ordered in decreasing FIR luminosity.  Condon (1992) has
found for a large sample of nearby spiral galaxies a mean value of $q$
= 2.3 with an rms scatter of 0.2.  The galaxies in our sample
generally fall within 2$\sigma$ of this value. 

The $z = 0$ values for $\alpha_{1.4}^{350}$ have a mean $\approx$ 0,
and an rms scatter ($\equiv 1\sigma$) of 0.14.  Using the polynomial
fits to the radio-to-IR SEDs, we have derived the evolution of
$\alpha_{1.4}^{350}$ with redshift for all 17 sources in Table 1.  The
results are shown in Figure 2. The solid curves show the distributions
for sources with L$_{\rm FIR} > 2\times10^{11}$ L$_\odot$, while the
dash curves show the  distributions
for sources with L$_{\rm FIR} < 2\times10^{11}$ L$_\odot$. 
The high luminosity sources have, on average, lower 
values of  $\alpha_{1.4}^{350}$ than the low luminosity sources.
The mean value of  $\alpha_{1.4}^{350}$ varies from 
$-0.026$ at $z = 0$  to +0.97 at $z = 7$ for the high luminosity
sources,  while the corresponding values for the
low luminosity sources are $-0.06$ and +1.18, respectively. 
Confirmation of this trend of $\alpha_{1.4}^{350}$ with FIR 
luminosity requires  significantly larger galaxy samples with well
observed radio-to-IR SEDs.

We derive the mean and rms scatter of $\alpha_{1.4}^{350}$ as a
function of redshift using the curves in Figure 2 for all 17 sources.
This mean-galaxy model is shown as the solid curve in Figure 3, while
the dotted curves delineate the $\pm 1 \sigma$ range, as
determined from scatter in the distribution of sources.   We 
designate these curves: $z_{mean}$, $z_{+}$, and $z_{-}$, respectively.
For a normal distribution, the $\pm 1 \sigma$ range implies that 67$\%$
of the sources in a given sample would fall within the range set by
the dotted curves in Figure 3.  The rms scatter is roughly constant
with redshift, decreasing from 0.14 at $z = 0$ to 0.12 at $z =
7$. Constant scatter in $\alpha_{1.4}^{350}$, combined with the
flattening of $\alpha_{1.4}^{350}$ with increasing $z$, implies that
the uncertainty in redshift estimates based on measured values of
$\alpha_{1.4}^{350}$ increases with $z$. Table 2 lists the mean, 
$z_{mean}$ (column 3), and $\pm$1$\sigma$ error ranges, $z_+$ and
$z_-$  (columns 2 and 4), for
likely redshifts given a measured value of $\alpha_{1.4}^{350}$
(column 1). 

We find that the mean galaxy model, $z_{mean}$, 
can be reasonably parameterized by a forth order polynomial: 
$ z = 0.050 - 0.308\alpha + 12.4\alpha^2 - 23.0\alpha^3 +
14.9\alpha^4 $, with a maximum deviation, $dz$, between the polynomial
fit and the mean galaxy model of $dz = 0.13$.  For values of
$\alpha_{1.4}^{350} \ge 0.9$, only a likely lower limit to a source
redshift can be derived.  The dashed curve in Figure 3 is the
$\alpha_{1.4}^{350}$ -- $z$ relationship derived for Arp 220, which
has been used as a representative template for the
$\alpha_{1.4}^{350}$ -- $z$ relationship by Barger, Cowie, \&
Richards (1999d).

The data points in Figure 3 are for observed submm sources with
measured spectroscopic redshifts, as summarized in Carilli \& Yun
(1999).  This includes a number of sources with AGN-type spectra, 
as discussed in detail in Yun et al. (1999). The sources H~1413+117 and
BR~0952$-$0115 are shown to demonstrate how the presence of a radio-loud
AGN affects the position of a source on this diagram. 
Carilli \& Yun (1999) point out that, given a source redshift, this
diagram can be used to search for evidence for a radio-loud AGN.


\subsection{Effects of Dust Temperature}

Blain (1999) has hypothesized that dust temperature may have an
important impact on the scatter in the $\alpha_{1.4}^{350}$ -- $z$
relationship, leading to a degeneracy between high redshift, hot dust
sources, and low redshift, cold dust sources. 
He showed that this effect can be mitigated by considering the 
quantity $(1 + z)\over{\rm T_D}$. In Figure 4 we plot the 
$\alpha_{1.4}^{350}$ values against $(1 + z)\over{\rm T_D}$,
using dust temperatures from Lisenfeld et al. (1999), G\"usten et
al. (1992), and Rigopoulou et al. (1996), with T$_{\rm D}$ 
normalized to the mean dust temperature of 39 K for the sample.
Again, the solid lines are for high luminosity sources, while the 
dash lines are for low luminosity sources. 

If we consider all the 
sources the rms is dominated by two sources, NGC 4418 and Mrk 231, 
and  the rms scatter of
$\alpha_{1.4}^{350}$ at a given value of  $(1 + z)\over{\rm T_D}$
has not decreased significantly from that in Figure 2 ($\sigma \approx
0.13$). However, if we remove the
two outliers we find a significant decrease in the
rms values, from 0.11 for the distribution in Figure 2, 
to 0.075 for the distribution in Figure 4, and the rms
is roughly constant with $(1 + z)\over{\rm T_D}$.

\subsection{Variations in Radio Properties}

The radio spectral indices from 330 MHz through 5 GHz for the sources 
listed in Table~1 show
a large scatter, from --0.2 to --0.9. The canonical value for
synchrotron emission from star forming galaxies is --0.8 (Condon
1992). Causes for flatter spectra include: (i) free-free absorption at 
low frequencies, (ii) free-free emission at high frequencies,
or (iii) a synchrotron 
self-absorbed AGN component.  Herein we examine the possible
contribution of free-free absorption to the observed values of
$\alpha_{1.4}^{350}$. 


We compare the observed spectral indices between 1.4 and 5 GHz to
those between 0.33 and 1.4 GHz  (columns 2 and 3 in Table 1). 
No evidence for systematic flattening of the spectra with 
increasing star forming activity is found
for the majority of the galaxies, as would
occur due to free-free absorption.  
However, the presence of 
a patchy absorbing medium could lead to complicated spectral shapes, 
and this alone is not a
sufficient reason to rule out the importance of free-free absorption.   

One method used by Condon et al. (1991) to investigate the possibility
of free-free absorption at 1.4 GHz was to correct the radio spectra by
assuming a constant radio spectral index of $-$0.8 and normalizing to
the observed flux densities at 5 GHz. This analysis will mitigate
free-free absorption effects at least out to $z = 2.6$ ($1 + z <
{5\over{1.4}}$).  We have investigated this possibility for the
galaxies in our sample with 5 GHz detections. We find that the rms
scatter at low redshift decreases only marginally, from 0.14 to 0.12.
Given these results, and the fact 
that free-free opacity decreases quadratically with increasing
frequency, we conclude that free-free absorption for rest 
frequencies above 
1.4 GHz is not likely to be a dominant cause of scatter in the
$\alpha_{1.4}^{350}$ -- $z$ relationship.
Multifrequency, high resolution radio imaging of a large sample of
star forming galaxies is required to properly address this issue.

\subsection{The Redshift Distribution of Faint Submm Sources}

Smail et al. (1999b) have discussed using  a statistical distribution
for source redshifts to provide constraints on models of the
star formation history of the universe using submm source counts.
Based on the original Carilli \& Yun (1999) and Blain (1999) models
for the $\alpha_{1.4}^{350}$ -- $z$ relationship,
they concluded that the median redshift for their faint submm sources
is likely to be between 2.5 and 3, with a conservative lower limit of
about 2. 

We have repeated the cumulative redshift distribution 
analysis of Smail et al. (1999b) using 
the values of $\alpha_{1.4}^{350}$ for the 14 field
galaxies in their Table 1, plus our new  
$\alpha_{1.4}^{350}$ -- $z$ model.
The results are shown in Figure 5. 
Note that for 7 of these galaxies only lower limits to 
$\alpha_{1.4}^{350}$ are available, which leads to lower limits to the
possible redshifts. We have performed the analysis using both 
$z_{\rm +}$ and $z_{\rm mean}$ curves, 
as shown in Figure 3. The $z_{\rm +}$ 
model leads to a conservative lower limit to the median redshift of
the sources of 2.05, while the $z_{\rm mean}$ model leads to a median
redshift of  2.68. 
These results strengthen the primary conclusions of Smail et
al. (1999b) that the majority of the faint submm sources are likely to
be at $z \ge 2$, and that there is no prominent low-z tail in the
distribution. 

There are three field galaxies in Table 1 of Smail et
al. (1999b) with reliable spectroscopic redshifts. For two of the
sources, SMMJ14011+0252 at $z_{\rm spec} = 2.55$ and SMMJ02399$-$0134
at $z_{\rm spec} = 1.06$, 
the redshifts predicted by the mean galaxy model in Figure 3
are remarkably close to the spectroscopic values,
$z_{\rm model}$ = 2.53 and  1.10, respectively.
For the source SMMJ02399$-$0136 the mean galaxy model 
predicts a redshift of 1.65, which is well below the spectroscopic
redshift of 2.80. This may indicate the presence of a radio-loud AGN
in this source, and the source has been shown to contain an optical
AGN, and possibly an extended radio jet source 
(Ivison et al. 1998). On the other hand, the source is 
not far beyond the limit of $z = 2.45$ predicted  by the 
$z_{\rm -}$ model in Figure 3.

For comparative illustration, the dotted line in Figure 5 shows the
expected fraction of uncollapsed 10$^{12}$ M$_\odot$ structures
derived using the standard CDM Press-Schechter formalism with a bias
factor 2 (Peebles 1993). There is an interesting, perhaps fortuitous,
similarity between the observed redshift distribution for the
formation of structures of this mass, comparable to elliptical
galaxies, and the redshift distribution of the faint submm sources. In
particular, structures of this mass first separate from the general
expansion of the universe and collapse under their own self-gravity at
$z \approx 5$, and are mostly in-place by $z \approx 1$ (Kaiser and
Cole 1989).

\section{Discussion}

We have derived the mean and rms scatter for redshift estimates based
on the $\alpha_{1.4}^{350}$ -- $z$ relationship from a sample of 17
low $z$ star forming galaxies with well sampled cm to infrared SEDs.
We emphasize that this analysis should be treated as strictly
statistical.  Given the rms scatter in $\alpha_{1.4}^{350}$, and the
shallow slope of the $\alpha_{1.4}^{350}$ -- $z$ relationship at high
redshift, the uncertainty in the redshift for any individual source is
large. However, for a large sample of sources, and assuming a normal
error distribution, one can say that 67$\%$ of the sources are within
the $\pm$1$\sigma$ curves in Figure 3, and that 84$\%$ of the sources
are at redshifts higher than the +1$\sigma$ curve in Figure 3. We have
repeated the analysis using the median values plus inner quartile
ranges, and reach essentially the same conclusions, although the
outliers tend to increase the error range when considering rms
deviations.

Applying the mean-galaxy model to the sample of Smail et al. (1999b),
we find that 80$\%$ of the sources are likely to be between $z = 1.5$
and 4. Blain et al. (1999a) have derived a cumulative source surface
density of 2.2 sources arcmin$^{-2}$ for source with flux densities
$\ge$ 1 mJy at 350 GHz, after correcting for magnification by
gravitational lensing. The implied comoving number density is then
$9.5 \times 10^{-4}$ Mpc$^{-3}$. These sources correspond to star
forming galaxies with FIR luminosities comparable to, or larger than,
that of Arp 220. The comoving space density of these high $z$ sources
is a factor 10$^3$ larger than that of low $z$ ultraluminous infrared
galaxies (Sanders and Mirabel 1996), and is comparable to that of low
$z$ elliptical galaxies (Lilly et al. 1999,
Cowie and Barger 1999). Hence the
normalization, and shape of the redshift distribution (see Figure 5),
for the faint submm sources are consistent with those
expected for forming elliptical galaxies (Tan et al. 1999).

The upper-most curve in Figures 2 and 4 is the distribution for NGC
4418, while the lowest curve is for Mrk 231.  Mrk 231 is well known to
contain a radio-loud AGN, although there is also evidence for radio
emission from a starburst disk (Downes and Solomon 1998, Carilli,
Wrobel, \& Ulvestad 1999a, Taylor et al. 1999). Below 1.4 GHz the
spectrum appears to be dominated by a possible star forming disk seen
on sub-kpc scales, while above 1.4 GHz the pc-scale AGN component
raises the total flux density by about a factor 2 (Taylor et
al. 1999). In fairness to high $z$ samples, for which such detailed
imaging information is not available, we have made no attempt to
remove the AGN emission for Mrk 231 in the mean galaxy model shown in
Figure 3.  The relatively flat radio spectrum of NGC 4418
($\alpha_{0.33}^{4.9} = -0.26$) may indicate that the source is
partially free-free absorbed, thereby leading to the source appearing
radio-quiet with respect to most low $z$ star forming galaxies. Also,
Lisenfeld et al. (1999) have pointed out that the dust spectrum of NGC
4418 is anomalous relative to the rest of their sample, with a low
value of $\beta$ and a high dust temperature, leading to the unusual
distribution for the source in the temperature-corrected curves shown
in Figure 4. Lisenfeld et al.  suggest that such an anomalous dust
spectrum may be due to significant dust opacity at submm
wavelengths. Whether such optical depth effects (free-free
absorption at cm wavelengths, and dust opacity at submm wavelengths)
are important for a significant fraction of high redshift sources
remains to be determined.

We have found that the scatter in
$\alpha_{1.4}^{350}$ decreases by about 30$\%$ when 
the dust temperature is also included in the analysis, 
as predicted by Blain (1999), with the
exception of the two anomalous sources discussed above. 
Unfortunately, even after correcting for dust temperature, the allowed
range in redshift for a given value of $\alpha_{1.4}^{350}$ is large,
eg. for $\alpha_{1.4}^{350}$ = 0.8 and T$_{\rm D}$ = 39 K the 
$\pm$1$\sigma$  redshift range is: 1.9 $ < z <$ 2.9, further
emphasizing that the $\alpha_{1.4}^{350}$ -- $z$ relationship 
should be used statistically, and not be considered an accurate
measure of an individual source's redshift. 
A further complication is that
accurate determination of T$_{\rm D}$ requires 
sensitive observations be made at a number of mm and submm
wavelengths. 

Implicit in the analysis of section 2 is the assumption
that there is not a systematic
change in the intrinsic SEDs for star forming galaxies with redshift,
as might occur due to cooling of the synchrotron emitting
relativistic electrons due to inverse Compton scattering off the
microwave background radiation at high $z$, or due to a systematic
change in ISM magnetic fields in high $z$ galaxies.  Such a variation
can only be tested with extensive observations at cm through submm
wavelengths of a large sample of high $z$ star forming galaxies with
known redshifts -- a task which remains problematic at the
present. The sparse data that currently exists
appear to be consistent with
the derived models based on low $z$ galaxies.

A massive star forming galaxy with a star formation rate of a few
hundred M$_\odot$ year$^{-1}$ at $z = 3$ has a submm flux density of a
few mJy at 350 GHz, and a radio flux density $\ge 20 \mu$Jy at 1.4 GHz
(Carilli \& Yun 1999, Cowie \& Barger 1999).  Such flux levels are
currently accessible with long integrations using existing bolometer
arrays on mm and submm telescopes, and using the Very Large Array
(VLA). In some studies  magnification by cluster gravitational lensing
has been used to improve the effective senstivities by a factor of 2
or so (Smail et al. 1997). Future instrumentation, such as the Atacama
Large Millimeter Array, and the upgraded VLA, should be able to detect
high $z$ galaxies with star formation rates of order 10 M$_\odot$
year$^{-1}$. Moreover, the very wide bandwidths ($\ge$ 8 GHz)
available on these future instruments will allow for searches for CO
emission over large redshift ranges in reasonable integration times,
thereby by-passing the requirement of optical spectroscopy for
accurate redshift determinations.

\vskip 0.2truein 

We would like to thank the referee, I. Smail, and R. Ivison, A. Blain
F. Owen, F. Bertoldi, and K. Menten, for useful comments and
discussions.  C.C. acknowledges support from the Alexander von
Humboldt Society.  The National Radio Astronomy Observatory is
operated by Associated Universities, Inc., under a cooperative
agreement with the National Science Foundation.  This research made
use of the NASA/IPAC Extragalactic Data Base (NED) which is operated
by the Jet propulsion Lab, Caltech, under contract with NASA.



\begin{center}
\begin{table}[htb]
\caption{Galaxy Sample}
\vskip 0.2in
\begin{tabular}{cccccc}
\hline
\hline
Source & $\alpha_{0.33}^{1.4}$ &  $\alpha_{1.4}^{4.9}$ &
 $\alpha_{1.4}^{350}$ & L$_{\rm FIR}$ & $q$ \\
~ & ~ & ~ & ~ & $\rm \times 10^{11} L_\odot$ & ~ \\
\hline
Mrk 231 & --0.48 &  +0.30 & --0.16 & 23  &  2.12 \\
Arp 220 & --0.15 & --0.37 &  +0.15 & 16  &  2.62 \\
Mrk 273 & --0.49 & --0.27 & --0.07 & 13  &  2.26 \\
UGC 5101 & --0.69 & --0.56 & --0.06 & 10  &  2.07 \\
NGC 6240 & --0.82 & --0.70 & --0.18 & 6.3  &  1.83 \\
Arp 193 & --0.54 & --0.52 & --0.01 & 4.3  &  2.33 \\
NGC 1614 &  --   & --0.54 & --0.03 & 3.6  &  2.50 \\
NGC 5256 & --0.79 & --0.67 & --0.09 & 2.8  &  1.95 \\
NGC 5135 & --0.73 & --0.50 & --0.07 & 1.7  &  2.08 \\
NGC 3110 & --0.65 & --0.79 & --0.01 & 1.6  &  2.16 \\
NGC 4418 & --0.27 & --0.26 &  +0.35 & 0.88  &  3.06 \\
NGC 5653 & --0.79 & --0.78 &  +0.05 & 0.84  &  2.38 \\
NGC 5936 &  --   & --0.89 & --0.14 & 0.82  &  1.95 \\
NGC 4194 & --0.56 & --0.77 &  +0.00 & 0.69  &  2.43 \\
M 82 & --0.34 & --0.61 & --0.11 & 0.48  &  2.20 \\
IRAS 05189-2524 &  --   & --0.50 &  0.18 & 10  &  2.76 \\
Zw 049 &  --   & --0.70 & --0.17 & 0.14 &  1.81 \\
\hline
\end{tabular}
\end{table}
\end{center}

\clearpage\newpage

\begin{center}
\begin{table}[htb]
\caption{Mean Galaxy Model}
\vskip 0.2in
\begin{tabular}{cccc}
\hline
\hline
$\alpha_{1.4}^{350}$ & $z_{\rm +}$ & $z_{\rm mean}$ & $z_{\rm -}$ \\
\hline
0.0 & 0 & 0 & 0.21 \\
0.1 & 0 & 0.14 & 0.40 \\
0.2 & 0.08 & 0.31 & 0.58 \\
0.3 & 0.23 & 0.49 & 0.81 \\
0.4 & 0.40 & 0.71 & 1.08 \\
0.5 & 0.60 & 0.96 & 1.42 \\
0.6 & 0.83 & 1.27 & 1.86 \\
0.7 & 1.12 & 1.65 & 2.45 \\
0.8 & 1.47 & 2.18 & 3.29 \\
0.9 & 1.93 & 2.95 & 4.44 \\
1.0 & 2.60 & 4.03 & 6.60 \\
1.1 & 3.63 & 6.04 & $>~7$ \\
\hline
\end{tabular}
\end{table}
\end{center}

\clearpage
\newpage

\begin{figure}
\psfig{figure=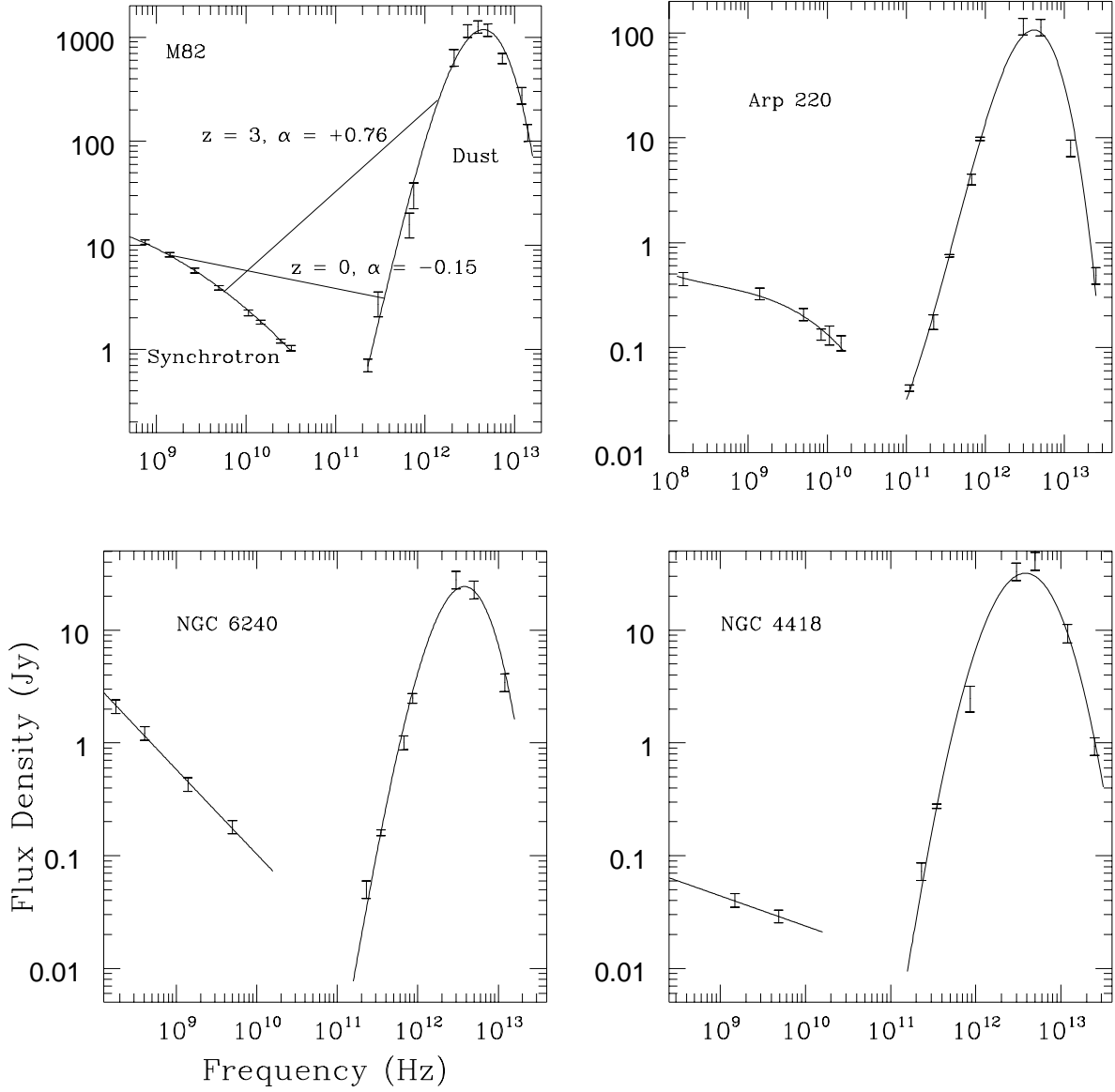,width=6.5in}
\caption{The data points are the radio through
infrared spectral energy distributions of 
four representative galaxies from our sample of
17 listed in Table 1. The data are obtained from the 
NED (the IRAS data points), the NVSS (Condon et al. 1998), the WENSS
(Rengelink et al. 1997), 
and frmo Rigopoulou, Lawrence, \&
Rowan-Robinson (1996), Benford (1999),and  Lisenfeld et al. (1999).
The solid curves are polynimial fits to the data. For the M82 
spectrum, the 
straight lines indicate the spectral index that would be derived  
for the source at $z = 0$ and at $z = 3$  between observing
frequencies of 1.4 and 350 GHz.
}
\end{figure}
\clearpage \newpage       

\begin{figure}
\psfig{figure=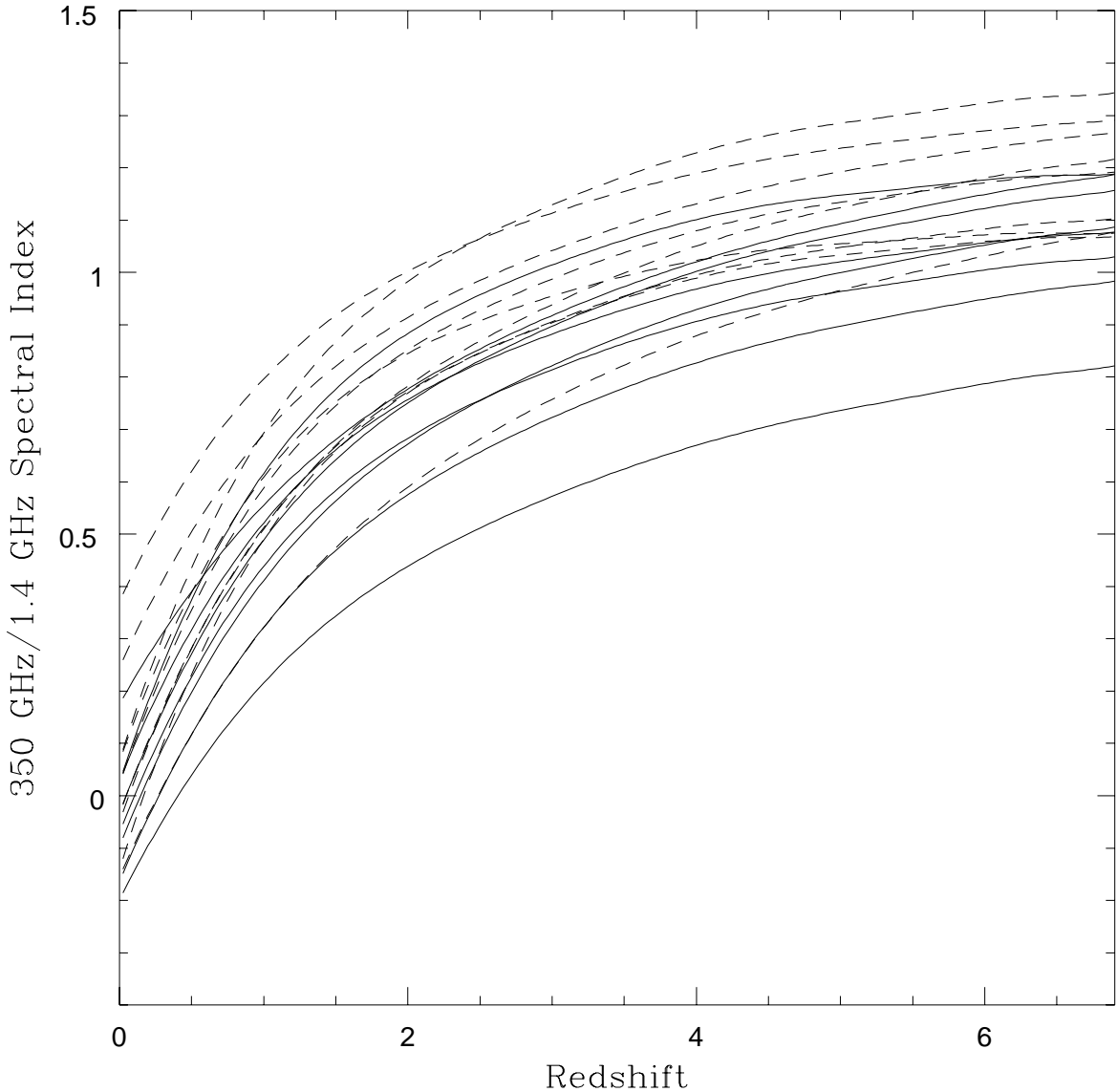,width=6.5in}
\caption{The   $\alpha_{1.4}^{350}$ -- $z$ relationship for the 17
galaxies in Table 1, derived from the polynomial fits to the cm
through IR SEDs.  The solid curves show the distributions
for sources with L$_{\rm FIR} > 2\times10^{11}$ L$_\odot$, while the
dashed curves show the  distributions
for sources with L$_{\rm FIR} < 2\times10^{11}$ L$_\odot$. 
Note that spectral index is related
to log flux density ratio by the scale factor: 
log[${350}\over{1.4}$] = 2.4. 
}
\end{figure}
\clearpage \newpage       

\begin{figure}
\psfig{figure=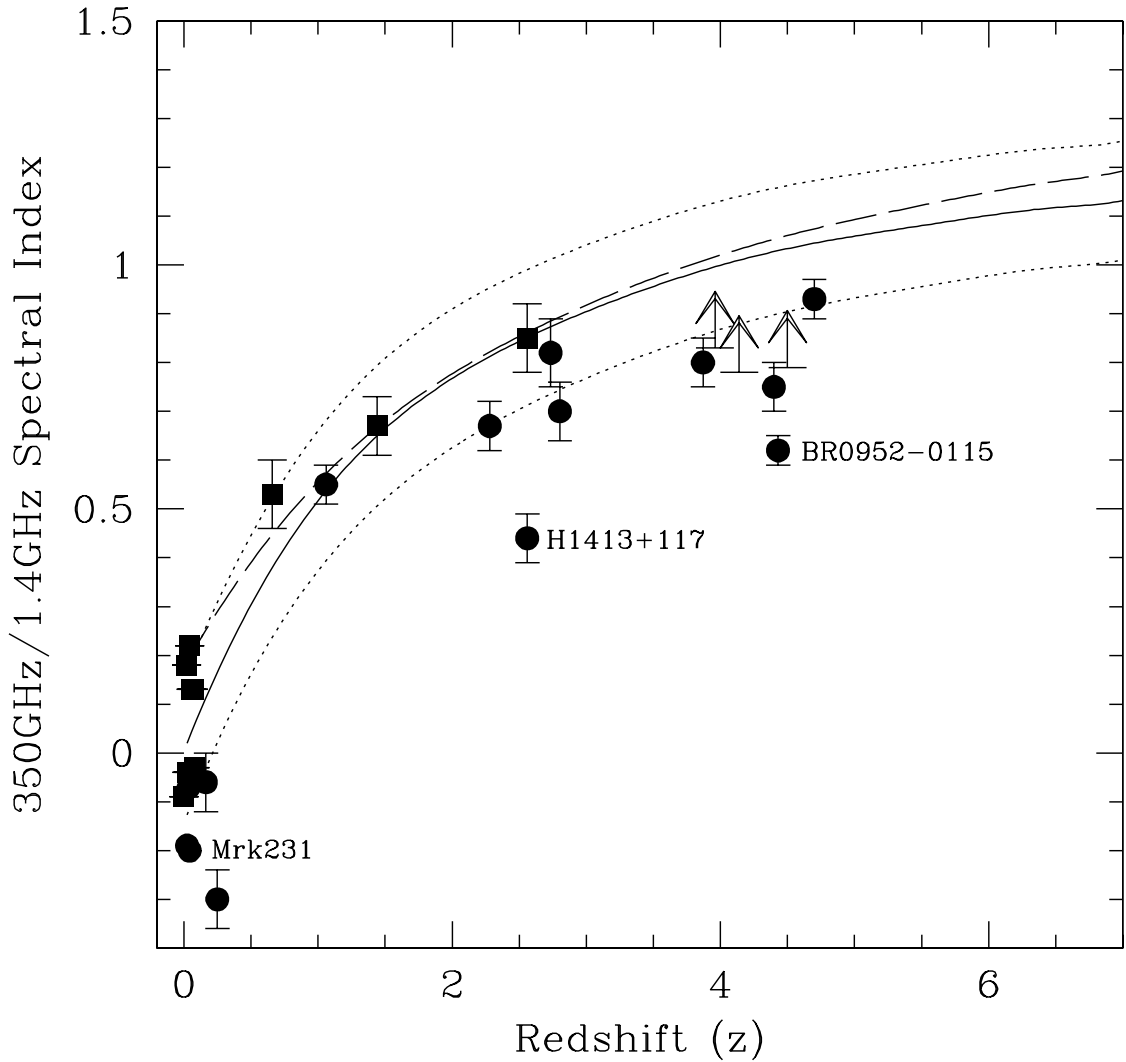,width=6.5in}
\caption{The mean $\alpha_{1.4}^{350}$ -- $z$ relationship for the 17
galaxies in Table 1, derived from the curves in Figure 3, is shown as
a solid line. The $\pm$1$\sigma$ curves are shown as dotted lines. 
In the text we  
designate these curves: $z_{mean}$ (solid), $z_{+}$ (dotted), and
$z_{-}$ (dotted), respectively.
The long dash line is for Arp 220.  Filled circles
(squares) represent galaxies with (without) an AGN signature
in optical spectrum.  A possible systematic vertical offset between
the galaxies with and without an AGN suggests  excess
radio emission associated with the AGN.    The data points are
for sources taken from Yun et al. (1999) and references therein.
}
\end{figure}
\clearpage \newpage

\begin{figure}
\psfig{figure=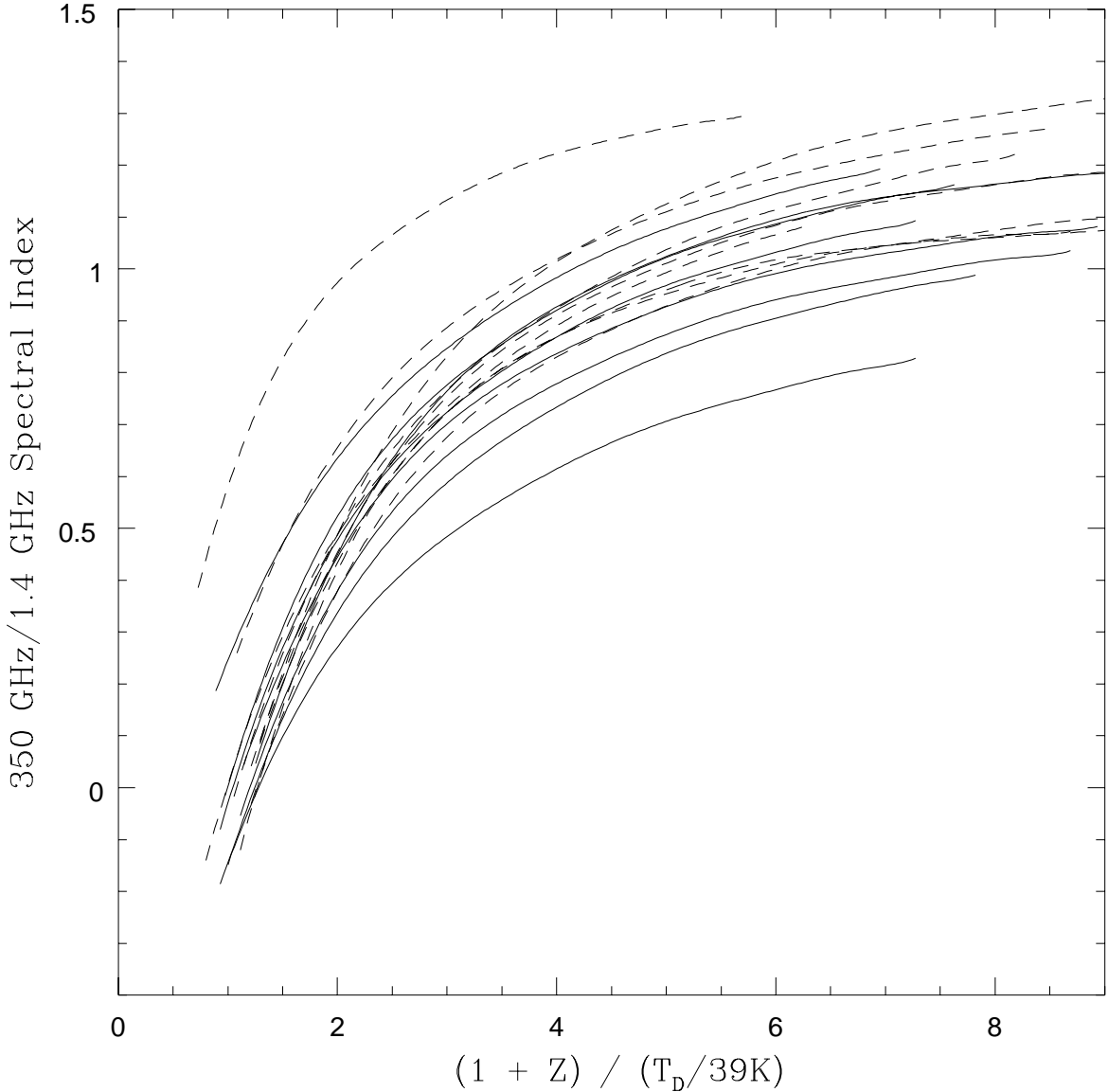,width=6.5in}
\caption{The   $\alpha_{1.4}^{350}$ -- $(1 + z)\over{\rm T_D}$ 
relationship for the 17
galaxies in Table 1, derived from the polynomial fits shown in Figure
2, and  normalized to the sample mean T$_{\rm D}$ = 39 K.
The solid curves show the distributions
for sources with L$_{\rm FIR} > 2\times10^{11}$ L$_\odot$, while the
dash curves show the  distributions
for sources with L$_{\rm FIR} < 2\times10^{11}$ L$_\odot$. 
}
\end{figure}
\clearpage \newpage       

\begin{figure}
\psfig{figure=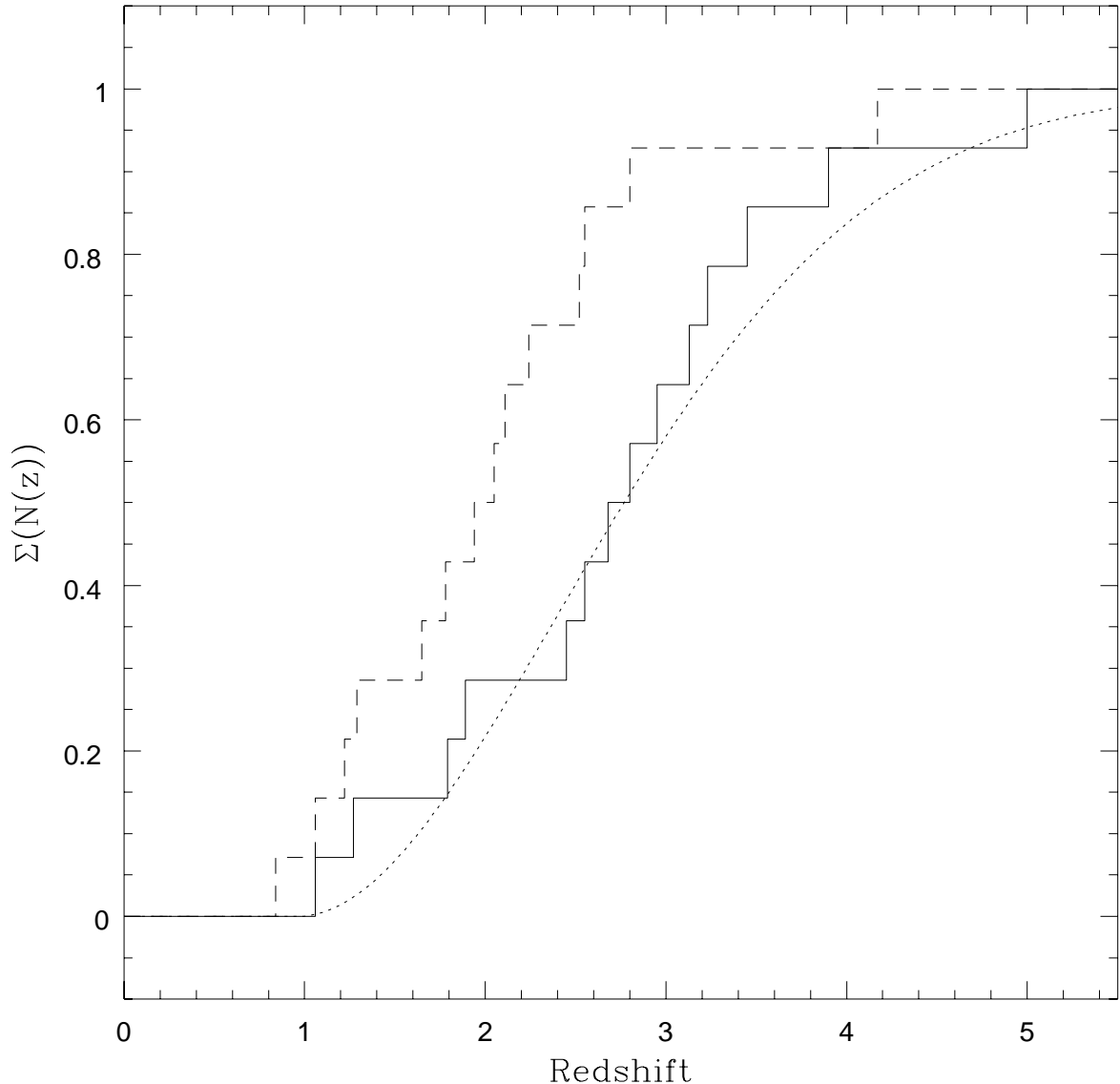,width=6.5in}
\vskip -0.4in
\caption{A plot of the cumulative redshift distribution for the 14
field galaxies from the Smail et al. (1999b) sample. The dash line is
the distribution predicted using the $z_{\rm +}$ model while the
solid line is the distribution predicted using the  $z_{\rm mean}$
model. Note that 7 of the sources in the Smail et al. (1999b)
sample have only lower limits to the values of $\alpha_{1.4}^{350}$, 
so these curves should be considered strictly lower
limits to the true distributions.  Also, for three of the galaxies 
reliable spectroscopic redshifts are available.
The dotted line shows the
expected fraction of uncollapsed 10$^{12}$ M$_\odot$ structures
derived using the standard CDM Press-Schechter formalism with a bias
factor 2 (Peebles 1993).
}
\end{figure}
\clearpage \newpage       
 
\centerline{\bf Erratum for: }
\centerline{\bf The Scatter in the Relationship between
Redshift and}
\centerline{\bf the Radio-to-Submm Spectral Index}

\vskip 0.5in

The 350~GHz flux densities of 4 of the 17 galaxies used in Carilli \&
Yun (2000) were found to be low by a factor 1.5 to 2 due to missing
extended emission in the original Lisenfeld et al. (2000) sample.
These values were subsequently up-dated by Lisenfeld et al. after our
paper went into press.  We have corrected these values using the
revised Lisenfeld et al.  data, which agree (within the errors) with
the data presented in Dunne, Clements, \& Eales (2000).  In addition,
two of the radio flux densities have been adjusted to correct for
confusing sources.

In this erratum we present a revised table, plus the revised model for
the relationship between redshift, $z$, and $\alpha_{1.4}^{350}$
$\equiv$ observed spectral index between 1.4~GHz and 350~GHz (Figure 6
below, Figure 3 in Carilli \& Yun (2000)).  The main change in the
model has been an increase in $\alpha_{1.4}^{350}$ by about 0.04 at
all redshifts (eg. from 0 to 0.04 at $z = 0$, and from 1.10 to 1.14 at
$z = 6$).  The rms scatter in  $\alpha_{1.4}^{350}$
remains roughly constant with redshift,
ranging from $\pm$0.17 at $z = 0$, to $\pm$0.14 at $z = 6$.  The
changes in the model are insufficient to change the basic conclusions
in Carilli \& Yun (2000).

In the revised figure we include the recent $\alpha_{1.4}^{350}$ --
$z$ model from Dunne et al. (2000), based on 104 low redshift
galaxies. The scatter in the Dunne et al.  model is about $\pm$ 0.10
in $\alpha_{1.4}^{350}$.  While there is substantial overlap between
the scatter ranges defined by the two distributions, the Dunne et
al. model predicts systematically higher values of
$\alpha_{1.4}^{350}$ at a given redshift.  For instance, at $z = 0$
the Dunne et al. model predicts $\alpha_{1.4}^{350}$ = 0.18$\pm$0.10,
while the 17 galaxy model in Figure 6 predicts $\alpha_{1.4}^{350}$ =
0.04$\pm$0.17.  This difference was pointed out by Dunne et al., based
on the model in Figure 3 in Carilli \& Yun (2000). They suggested that
much of the difference might be due to the low submm flux densities
used for a few of the sources in the original model. The revised 17
galaxy model presented herein shows that about 30$\%$ of the
difference can be explained by the errant data.  If we also remove the
two galaxies with evidence for a radio AGN (Mrk 231 and NGC 6240), the
$z = 0$ value of $\alpha_{1.4}^{350}$ rises to 0.08. The remaining
difference could be due to the fact that most of the sources in our 17
galaxy sample are in the upper half of the luminosity distribution
delineated by the 104 galaxies used by Dunne et al. (ie. L$_{\rm 1.4
GHz}$ $\ge$ $5 \times 10^{22}$~W~Hz$^{-1}$). Figure 3 in Dunne et
al. shows a systematic decrease in $\alpha_{1.4}^{350}$ with
increasing radio luminosity (a fact also pointed out in Carilli \& Yun
2000), from about +0.3 for galaxies with radio spectral luminosities
at 1.4~GHz of about $3 \times 10^{21}$~W~Hz$^{-1}$, to +0.05 for
galaxies with radio luminosities of $3 \times 10^{23}$~W~Hz$^{-1}$. On
the other hand, the remaining difference is well within the scatters
of the two distributions, and could simply be due to limited
statistics.

\vskip 0.3in

\centerline{\bf References}

Dunne, L., Clements, D., and Eales, S. 2000, MNRAS, in press
(astro-ph/0002436)

Carilli, C.L. and Yun, M.S. 2000, ApJ, 530, 618

Lisenfeld, U., Isaak, K.,G., and Hills, R. 2000, MNRAS (letters), 312,
433 

\begin{center}
\begin{table}[htb]
\caption{Galaxy Sample}
\vskip 0.2in
\begin{tabular}{ccccccc}
\hline
\hline
Source & $\alpha_{0.33}^{1.4}$ &  $\alpha_{1.4}^{4.9}$ &
 $\alpha_{1.4}^{350}$ & L$_{\rm FIR}$ & L$_{\rm 1.4~GHz}$ & $q$ \\
~ & ~ & ~ & ~ & $\times 10^{11}$ L$_\odot$ & $\times 10^{23}$ W
Hz$^{-1}$  & ~ \\
\hline
Mrk 231 & --0.48 &  +0.30 & --0.16 & 23  & 10.0 & 2.12 \\
Arp 220 & --0.15 & --0.37 &  +0.15 & 16  & 2.0 &  2.62  \\
Mrk 273 & --0.49 & --0.27 & --0.07 & 13  & 4.0 &  2.26  \\
UGC 5101 & --0.69 & --0.56 & --0.06 & 10  & 4.8 &  2.07  \\
IRAS 05189-2524 &  -- & --0.50 & +0.18 & 10  &  1.0 & 2.76 \\
NGC 6240 & --0.82 & --0.70 & --0.18 & 6.3  & 5.0 & 1.83 \\
Arp 193 & --0.54 & --0.52 & --0.01 & 4.3  &  1.0 & 2.33 \\
NGC 1614 &  --   & --0.54 & +0.10 & 3.6  &  0.61 & 2.50  \\
NGC 5256 & --0.79 & --0.67 & --0.09 & 2.8  &  1.7 & 1.95  \\
NGC 5135 & --0.73 & --0.50 & --0.07 & 1.7  &  0.73 & 2.08  \\
Zw 049 &  -- & -- &      +0.24 & 1.7 &  0.18 & 2.71  \\
NGC 3110 & --0.65 & --0.79 & +0.10 & 1.6  &  0.62 & 2.16  \\
NGC 4418 & --0.27 & --0.26 &  +0.35 & 0.88  &  0.04 & 3.06   \\
NGC 5653 & --0.79 & --0.78 &  +0.20 & 0.84  &  0.19 & 2.38  \\
NGC 5936 &  --   & -- &  +0.20 & 0.82  &  0.16 & 2.40   \\
NGC 4194 & --0.56 & --0.77 &  +0.00 & 0.69  &  0.14 & 2.43  \\
M 82 & --0.34 & --0.61 &    --0.11 & 0.48  &  0.09 & 2.20  \\
\hline
\end{tabular}
\end{table}
\end{center}

\clearpage\newpage

\begin{figure}
\psfig{figure=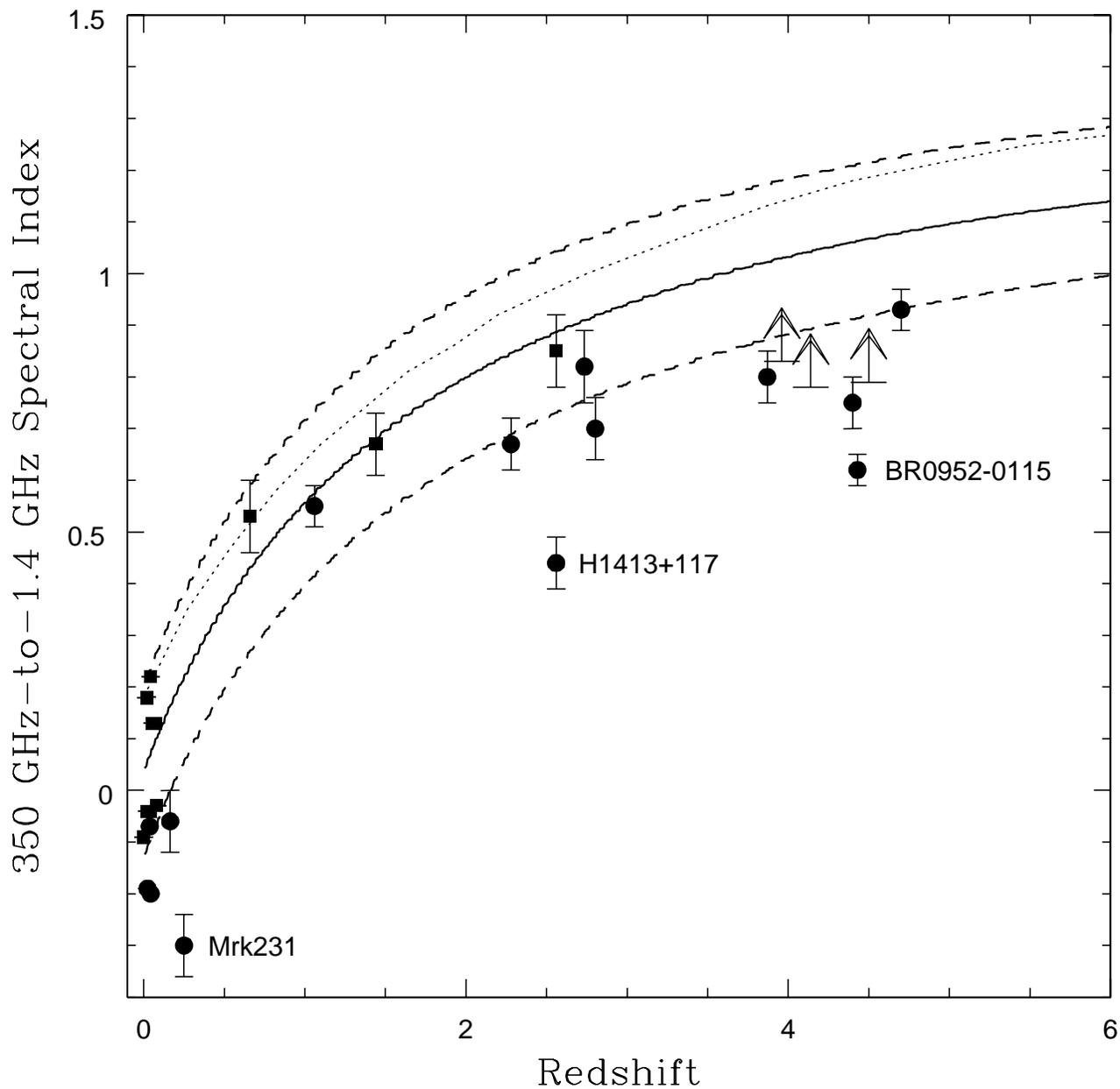,width=7in}
\caption{The mean $\alpha_{1.4}^{350}$ -- $z$ relationship for the 17
galaxies in Table 1 is shown as
a solid line (originally Figure 3 in Carilli and Yun (2000)). 
The $\pm$1$\sigma$ curves are shown as dashed lines.
The  dotted line is the $\alpha_{1.4}^{350}$ -- $z$ relationship for 
104 from Dunne et al. (2000).  The data points are the same
as in Figure 3 in Carilli and Yun (2000).
}
\end{figure}
\clearpage \newpage

\end{document}